\documentclass[superscriptaddress, reprint, amsmath, amssymb, aps, prl]{revtex4-1}

\usepackage{graphicx}%
\usepackage{xspace}%

\newcommand{\ie}{\emph{i.e.}\xspace}%
\newcommand{\muSR}{$\mu$SR\xspace}%
\newcommand{\lem}{LE-$\mu$SR\xspace}%
\newcommand{\SL}[1]{[3LSCO+#1LCO]\xspace}%

\begin{document}

\title{Superconductivity in $\mathrm{La_{1.56}Sr_{0.44}CuO_4}/\mathrm{La_2CuO_4}$ superlattices}

\author{A. Suter}%
\email{andreas.suter@psi.ch}%
\affiliation{Laboratory for Muon Spin Spectroscopy, Paul Scherrer Institute, 5232 Villigen PSI, Switzerland}%
\author{E. Morenzoni}%
\affiliation{Laboratory for Muon Spin Spectroscopy, Paul Scherrer Institute, 5232 Villigen PSI, Switzerland}%
\author{T. Prokscha}%
\affiliation{Laboratory for Muon Spin Spectroscopy, Paul Scherrer Institute, 5232 Villigen PSI, Switzerland}%
\author{B. M. Wojek}%
\affiliation{Physik-Institut, Universit\"{a}t Z\"{u}rich, 8057 Z\"{u}rich, Switzerland}%
\affiliation{Laboratory for Muon Spin Spectroscopy, Paul Scherrer Institute, 5232 Villigen PSI, Switzerland}%
\author{H.~Luetkens}%
\affiliation{Laboratory for Muon Spin Spectroscopy, Paul Scherrer Institute, 5232 Villigen PSI, Switzerland}%
\author{G. Nieuwenhuys}%
\affiliation{Laboratory for Muon Spin Spectroscopy, Paul Scherrer Institute, 5232 Villigen PSI, Switzerland}%
\author{A. Gozar}%
\affiliation{Brookhaven National Laboratory, Upton, New York 11973-5000, USA}%
\author{G. Logvenov}%
\altaffiliation{present address: Max-Planck-Institut f\"{u}r Festk\"{o}rperforschung, 70569 Stuttgart, Germany}%
\affiliation{Brookhaven National Laboratory, Upton, New York 11973-5000, USA}%
\author{I. Bo\v{z}ovi\'{c}}%
\affiliation{Brookhaven National Laboratory, Upton, New York 11973-5000, USA}%
\date{\today}%

\begin{abstract}
Superlattices of the repeated structure $\mathrm{La_{1.56}Sr_{0.44}CuO_4}/\mathrm{La_2CuO_4}$ (LSCO-LCO), where none of the constituents is superconducting, show a superconducting transition of $T_{\rm c} \simeq 25$ K. In order to elucidate the nature of the superconducting state we have performed a low-energy \muSR study. By applying a magnetic field parallel (Meissner state) and perpendicular (vortex state) to the film planes, we could show that superconductivity is sheet like, resulting in a very anisotropic superconducting state. This result is consistent with a simple charge-transfer model, which takes into account the layered structure and the difference in the chemical potential between LCO and LSCO, as well as Sr interdiffusion. Using a pancake-vortex model we could estimate a strict upper limit of the London penetration depth to 380 nm in these superlattices. The temperature dependence of the muon depolarization rate in field cooling experiments is very similar to what is observed in intercalated BSCCO and suggests that vortex-vortex interaction is dominated by electromagnetic coupling but negligible Josephson interaction.
\end{abstract}

\pacs{76.75.+i, 74.20.-z, 74.25.Ha, 74.25.Nf, 74.78.-w}%
\keywords{low dimensional magnetism, interface superconductivity}%
\maketitle

In thin interfacial layers inside oxide heterostructures a host of electronic states were discovered experimentally, as for instance a high-mobility 2D electron gas \cite{ohtomo_high-mobility_2004}, magnetism \cite{brinkman_magnetic_2007}, quantum Hall effect \cite{tsukazaki_quantum_2007}, and interface superconductivity between insulators \cite{reyren_superconducting_2007}. In metal-insulator $\mathrm{La_{1.56}Sr_{0.44}CuO_4}/\mathrm{La_2CuO_4}$ (LSCO-LCO) heterostructures, where none of the constituents is superconducting, superconductivity with a $T_{\mathrm{c}} \approx 40$ K has been discovered recently \cite{gozar_high-temperature_2008}. Subsequent experiments on LSCO-LCO bi-layers, deploying Zn doping in individual layers, have established that superconductivity is present at the interfaces only \cite{logvenov_high-temperature_2009}, and that Sr interdiffusion is limited to about 1 unit cell (UC) \cite{gozar_high-temperature_2008}. The physics of LSCO-LCO superlattices (SLs) is very interesting due to the close proximity of superconductivity and magnetism in these systems, and the potentially very weakly coupled superconducting layers. 

Here we present an investigation about the superconducting properties of $\mathrm{La_{1.56}Sr_{0.44}CuO_4}/\mathrm{La_2CuO_4}$ SLs. Counting in $1/2$-UC increments, each of which contains a single $\mathrm{CuO_2}$ plane, the investigated SLs have the repeated structure \SL{6}, \SL{9}, and \SL{12}. All SLs were $c$-axis oriented grown on $\mathrm{LaSrAlO_4}$ substrates. The total film thickness was kept to about 85 nm. Mutual induction measurements show that these SLs have a superconducting transition at $T_{\rm c} \simeq 25$ K. Further details about the sample growth and characterization can be found in Refs. \cite{gozar_high-temperature_2008,logvenov_high-temperature_2009,suter_mag_SL_20011}. In Ref. \cite{suter_mag_SL_20011} we have shown that LCO within these SLs is indeed magnetic and spontaneous zero field precession is found in \SL{12}. In all the SLs discussed here LCO is magnetic even though the magnetic state is extremely soft (low spin stiffness) compared to bulk LCO and already in the \SL{9} SL quantum fluctuations are substantially enhanced such that no zero field precession is observable anymore.

\begin{figure}[ht!]
 \centering%
 \includegraphics[width=0.35\textwidth]{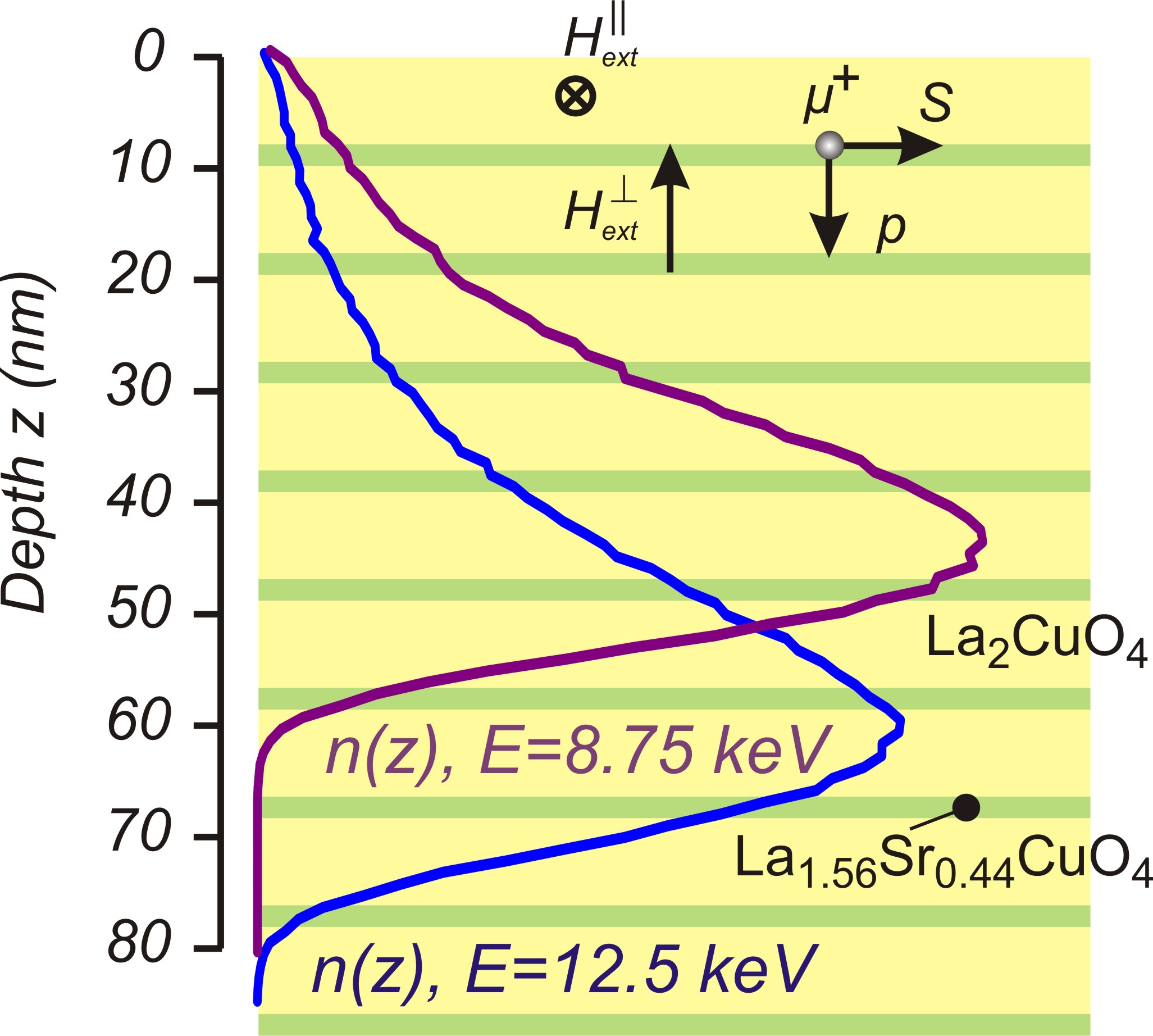}\qquad%
 \begin{minipage}[b]{9.2cm}
 \caption{$\mu^+$ stopping distribution, $n(z)$, for the \SL{12} superlattice. The yellow stripes represent the
          LCO, the green ones the LSCO.
          In the Meissner screening experiments, after zero field cooling, an external magnetic field, 
          $H_{\rm ext}^{\|}$, parallel to the layers ($\|$ to the $ab$-planes) is applied. These Meissner screening
          experiments were performed at a $\mu^+$ implantation energy, $E_{\rm impl}=8.75$ keV. At this energy, any
          Meissner screening would lead to a maximal diamagnetic shift of the $\mu^+$ precession signal.
          For the vortex lattice experiments a field $H_{\rm ext}^{\perp}$ perpendicular to the layers ($\|$ to 
          the $c$-axis) was applied, and an implantation energy of $E_{\rm impl}=12.5$ keV was chosen to maximize 
          the coverage of the SL.
 }\label{fig:LSCO-SL-TrimSP}
 \end{minipage}
\end{figure}

The microscopic investigation of the superconducting state was carried out with two sets of experiments by means of \lem, namely in the Meissner screening geometry ($H_{\rm ext}^\|$, \ie field parallel to the film) and in the vortex state ($H_{\rm ext}^\perp$, \ie field perpendicular to the film), both are $\mu^+$ transverse field geometries as depicted in Fig. \ref{fig:LSCO-SL-TrimSP}, where also the $\mu^+$ stopping distributions, $n(z)$ for the chosen experiments are shown. We will start the discussion with the Meissner screening. Here the samples are zero field cooled and afterwards an external magnetic field, $H_{\rm ext}^{\|} < H_{\rm c1}^{\|} \simeq \sqrt{24}\, (\lambda/t)\, H_{\rm c1}$ (see Ref.\cite{tinkham_introduction_2004}), parallel to the SL is applied, where $H_{\rm c1}$ is the bulk lower critical field, $\lambda$ the magnetic penetration depth, and $t \simeq 85$ nm the sample thickness. The chosen implantation energy $E_{\rm impl}=8.75$ keV having its peak in the center of the SLs, hence if there would be any substantial Meissner screening present, the measured muon precession frequency would show a diamagnetic shift. In the experiment we applied a field of $\mu_0\, H_{\rm ext}^{\|} = 10$ mT. Down to $T=5$ K no diamagnetic shift was found within the experimental resolution. This shows there is no substantial amount of supercurrents flowing perpendicular to the SLs, indicating that superconductivity in these SLs must be sheet like, and strengthening the conclusions form previous transport measurements \cite{bozovic_no_2003}.

\begin{figure}[ht!]
 \centering%
 \includegraphics[width=0.45\textwidth]{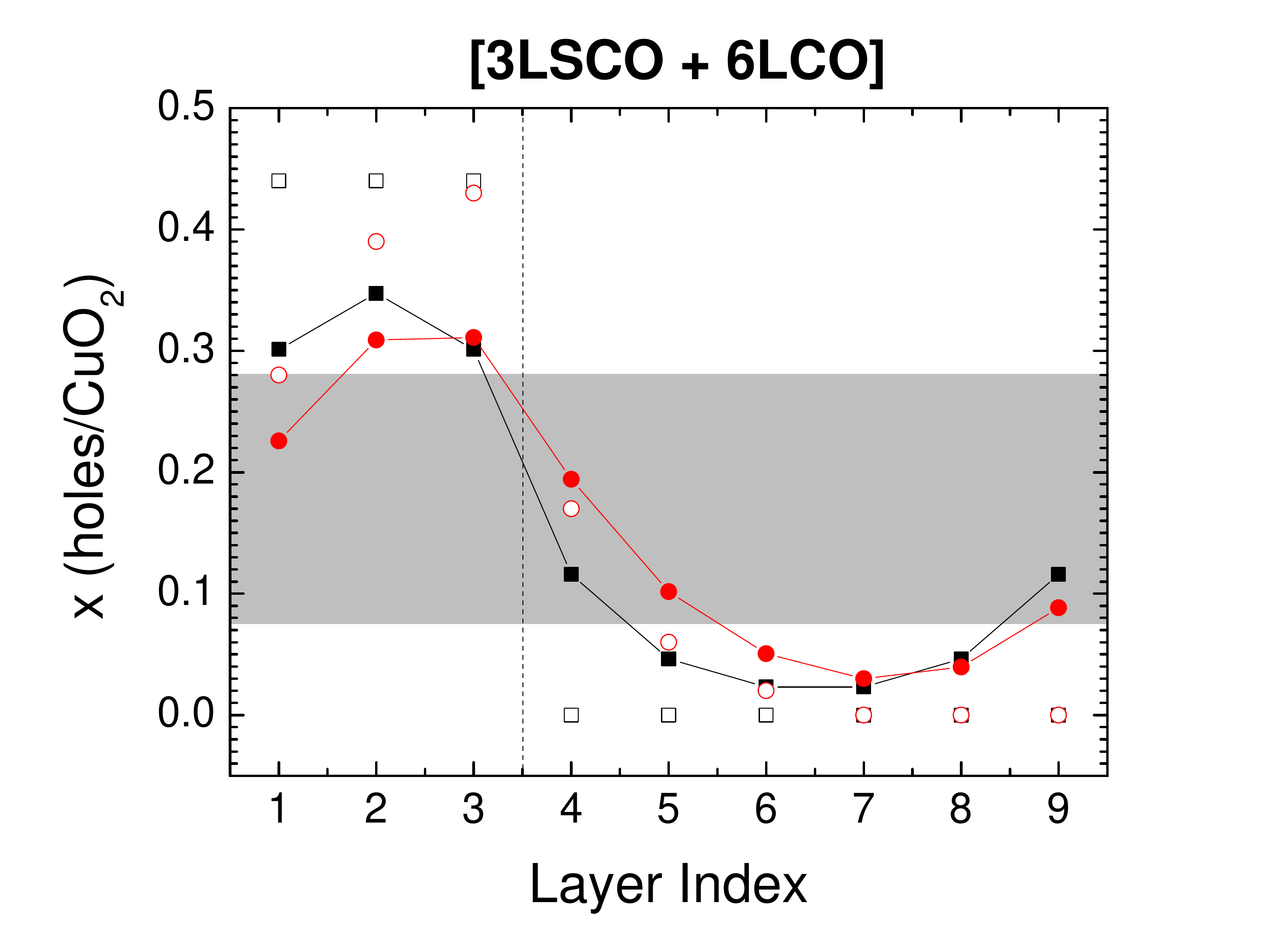}\qquad%
 \includegraphics[width=0.45\textwidth]{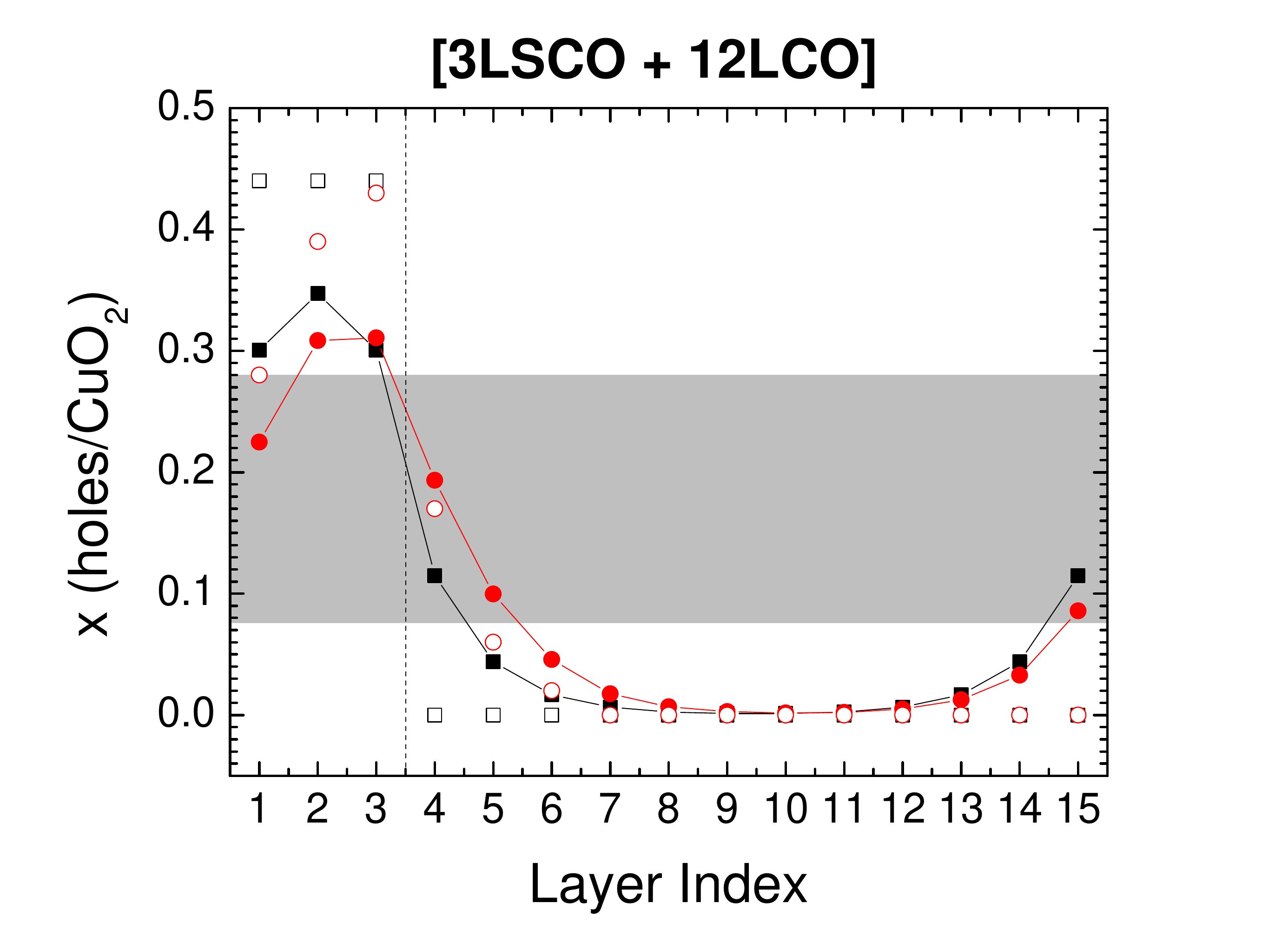}%
 \caption{Calculated charge profiles for the \SL{6} and \SL{12} SLs, respectively. The gray band shows the doping range 
          for which superconductivity takes place in bulk LSCO. The open symbols show the nominal charge levels in 
          the SLs. Open black squares: nominal charge level assuming no Sr interdiffusion; open red circles: nominal
          charge level with Sr interdiffusion taken into account. The solid symbols show the charge distribution
          calculated allowing for charge-transfer. Solid black square: assuming no Sr interdiffusion; 
          solid red circles: with Sr interdiffusion taken into account.}\label{fig:LSCO-SL-ChargeProfiles}
\end{figure}

\noindent The sheet-like nature of the superconducting state can be quantified using a simple charge-transfer model. The charge modulation within the SL can be estimated from an approach similar to the one described in Ref. \cite{loktev_model_2008}. Starting from a discrete version of Poisson's equation

\begin{equation}\label{eq:poisson}
 \Delta \phi \simeq \phi_{j+1} + \phi_{j-1} - 2 \phi_j = 
  -\underbrace{\frac{e c}{\epsilon_0\epsilon_{\rm r} a^2}}_{\displaystyle \equiv \alpha}\, \delta_j
\end{equation}

\noindent where $\phi_j$ is the potential in layer $j$, $e$ is the elementary charge, $\epsilon_0$ the dielectric constant, and $\epsilon_{\rm r}$ the relative dielectric constant of the solid. $a$ and $c$ are the in-plane and out-of-plane lattice constants. $\delta_j$ is the induced number of holes per Cu plaquette. The chemical potential $\mu_j$ is driving the charge transfer and satisfying the equation $\mu_j + e \phi_j = \mathrm{const}$. Linearizing locally the chemical potential results in

\begin{equation}\label{eq:linearized_chemical_potential}
 \phi_{j+1} - \phi_j = -\frac{1}{e} (\mu_{j+1}-\mu_j) \simeq -\underbrace{\frac{1}{e} 
 \frac{\partial\mu}{\partial n}}_{\displaystyle \equiv \beta} (n_{j+1}-n_j)
\end{equation}

\noindent where $n_j = x_j + \delta_j$, and $x_j$ is the nominal doping level due to Sr. Combining Eqs. (\ref{eq:poisson}) and (\ref{eq:linearized_chemical_potential}), one can eliminate $\phi_j$, resulting in

\begin{eqnarray}\label{eq:charge_distribution}
 -(1+\gamma) \delta_1 + \delta_2 &=& x_1 - x_2 \nonumber \\
 \delta_{j-1} - (2+\gamma)\delta_j + \delta_{j+1} &=& -x_{j-1} + 2 x_{j} - x_{j+1}, \quad \forall j \in ]1, n[ \\
 \delta_{n-1} -(1+\gamma) \delta_n &=& -x_{n-1} + x_n \nonumber 
\end{eqnarray}

\noindent where $\gamma = \alpha / \beta$. Figure \ref{fig:LSCO-SL-ChargeProfiles} shows the charge profiles obtain by solving the set of equations (\ref{eq:charge_distribution}), for $\epsilon_{\rm r} = 30$, and the Thomas-Fermi screening length of $\lambda_{\rm TF}^2 = [6\,\mathrm{\AA}]^2 = \epsilon_0/e^2 (\partial\mu/\partial n) = \epsilon_0/e \beta$, corresponding to experimental results for $\partial\mu/\partial n$ \cite{ino_chemical_1997}, as well as the value for $\lambda_{\rm TF}$ estimated from resonant soft X-ray scattering experiments \cite{smadici_superconducting_2009}.

\noindent The gray band in Fig. \ref{fig:LSCO-SL-ChargeProfiles} indicates the nominal doping region where superconductivity occurs. This indicates, since Sr interdiffusion is small, that the charge transfer is only slightly smeared out compared to the ideal situation. It also shows that in these SLs superconducting sheets form in LCO next to the interface with LSCO, which reduces the number of magnetic layers within the LCO slabs. These estimates explain naturally the absence of Meissner screening and are consistent with our findings on the magnetic properties in the \SL{12} SL \cite{suter_mag_SL_20011}. The typical distance between these superconducting sheets is about 1 UC = 13.2 \AA~ between the two LSCO interfaces, and $d \approx (N/2-1.5)$ UC between the LCO layers ($N$ is the $1/2$-UC counting for the LCO, \emph{e.g.} $d \approx 60$ \AA~ for the \SL{12} SL for which $N=12$). These are huge distances to couple the superconducting sheets to acquire long-range phase coherence necessary to drive the superconducting transition observed in the experiment. This is somewhat similar to the situation in [Bi-2212 $+\, N\times$Bi-2201] SLs \cite{bozovic_sperconductivity_in_BSCCO_SLs_92} and in Bi-2212 and Bi-2201 intercalated with organic molecules \cite{baker_tuninginterlayer_2009}, except that in the SL case discussed here the distance $d$ is about ten times larger. Therefore, we expect Josephson coupling to be negligibly small, resulting in a pancake-vortex state \cite{clem_two-dimensional_1991} where the vortex pancake-pancake interaction is reduced to mere dipolar interaction. 

\begin{figure}[ht!]
 \centering%
 \includegraphics[width=0.42\textwidth]{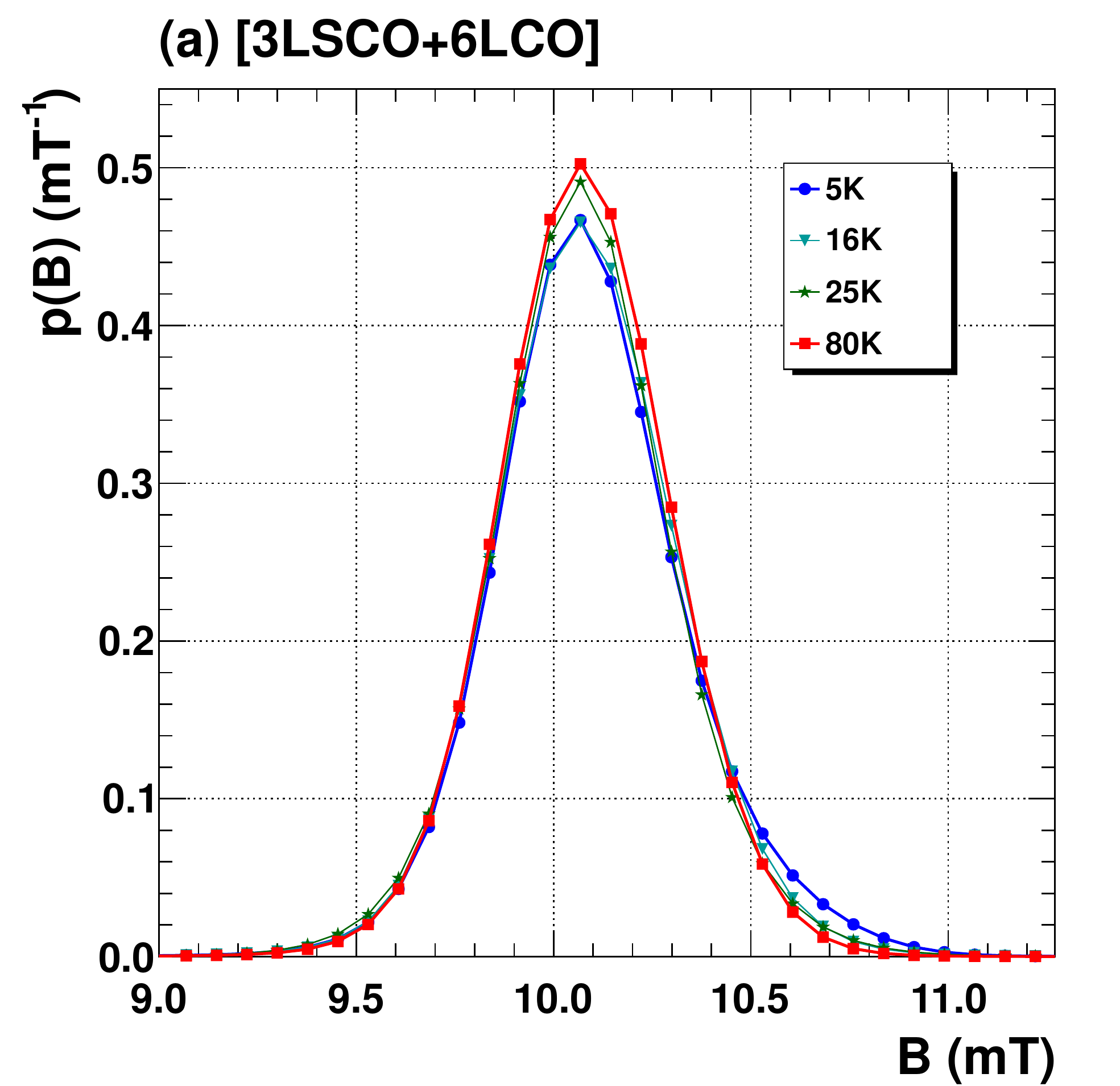}\qquad%
 \includegraphics[width=0.42\textwidth]{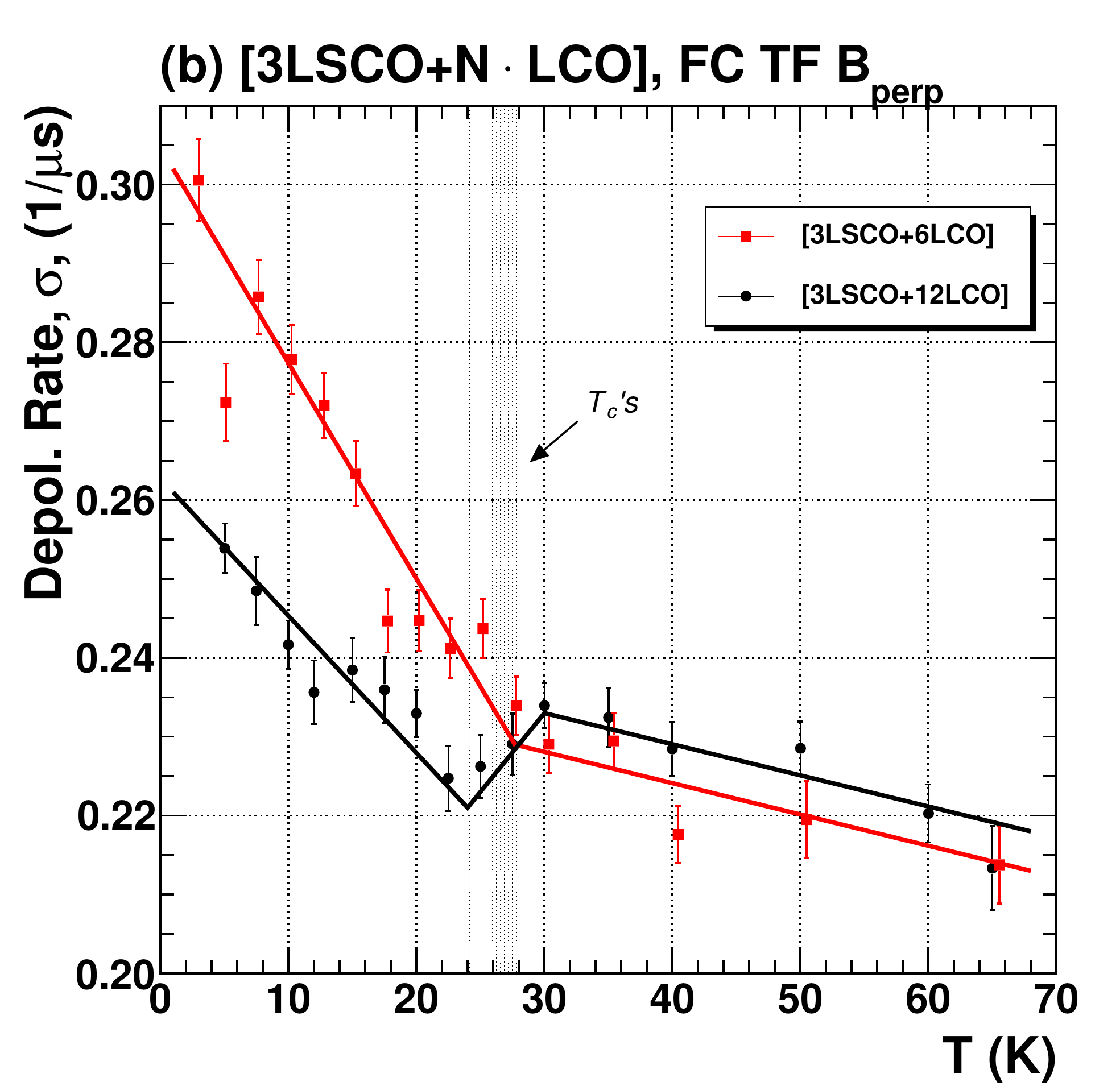}%
 \caption{(a) Normalized magnetic field distribution for the \SL{6} superlattice for $E_{\rm impl}=12.5$ keV. 
          (b) Transverse muon depolarization rate versus temperature for the same implantation energy. 
          The SLs were FC in a field of $\mu_0 H_{\rm ext}^\perp=10.1$ mT. 
          The lines are guides to the eye.}\label{fig:LSCO1to3_MaxEnt_and_LSCO-SL-FC-TFrate}
\end{figure}

To investigate the vortex state of the SLs we carried out experiments in the $H_{\rm ext}^\perp$ geometry (see Fig.  \ref{fig:LSCO-SL-TrimSP}) while field cooling the SLs, and choosing an implantation energy of $E_{\rm impl} = 12.5$ keV. Fig. \ref{fig:LSCO1to3_MaxEnt_and_LSCO-SL-FC-TFrate}a shows the normalized magnetic field distribution, $p(B)$, obtained by maximum entropy analysis \cite{riseman_maxent_2000} for \SL{6}, of measurements carried out at $\mu_0 H_{\rm ext}^\perp = 10.2$ mT. On the high field side, a shoulder can be seen typical for the formation of a vortex lattice, with weight that increases when cooling down through the superconducting transition. The effect is rather small, as also apparent from Fig. \ref{fig:LSCO1to3_MaxEnt_and_LSCO-SL-FC-TFrate}b where the Gaussian depolarization rate, $\sigma$, versus temperature is shown. The temperature dependence of $\sigma$ is very atypical for vortex broadening measured by \muSR which usually follows a functional form $\sigma \propto [ 1 - (T/T_{\rm c})^r ]$, with $r\simeq 2\ldots 6$ \cite{baker_tuninginterlayer_2009}. However, the behavior found here is very similar to the one found for highly intercalated BSCCO samples \cite{baker_tuninginterlayer_2009}. The authors interpreted this as the breakdown of the Josephson coupling between the vortices. This is even more likely for the SLs as already mentioned when estimating the charge-transfer effects. Still we will try to relate the increase of $\sigma$ below $T_{\rm c}$ to the London penetration depth $\lambda_{\rm L}$. According to Ref. \cite{brandt_properties_2003} the in-plane magnetic screening length, $\lambda_\|$ is related to the superconducting vortex broadening of the \muSR signal as

\begin{equation}
 \sigma_{\rm sc}^2(T) = \sigma^2(T<T_{\rm c}) - \sigma^2(T_{\rm c}) = 0.00371 \times 
    \left[ \frac{\gamma_\mu \Phi_0}{\lambda_\|^2(T)} \right]^2,
\end{equation}

\noindent where $\Phi_0 = h/(2e)$ is the flux quantum, $h$ the Planck constant, and $e$ the electron charge. For a layered superconductor, the London penetration depth $\lambda_{\rm L}(T)$, which is a measure of the superfluid density $n_{\rm S} \propto 1/\lambda_{\rm L}^2$, is given as $\lambda_{\rm L} = (d_{\rm sc}/s) \cdot \lambda_\|$ \cite{clem_two-dimensional_1991}, where $d_{\rm sc}$ is the thickness of the superconducting layer, and $s$ the spacing distance between the superconducting sheets. Taking all these considerations into account, one arrives at an estimate of $\lambda_{\rm L} \simeq 380\,\mathrm{nm} \approx 1.5 \times \lambda_{\rm L}^{\rm bulk,opt.}$ (for $\lambda_{\rm L}^{\rm bulk,opt.}$ see Ref. \cite{luke_bulk_lsco_1997}). We would like to stress that this is a strict \emph{upper} limit for $\lambda_{\rm L}$, since the theoretical framework relating $\sigma_{\rm sc}$ to $\lambda_{\rm L}$ assumes Josephson coupling to be present, as well as the presence of a regular vortex lattice. Neither is likely to be case in the investigated SLs. Furthermore $\sigma_{\rm sc}$ is diminished in thin films due to widening of the flux lines close to the surface \cite{niedermayer_direct_observation_99}.

In conclusion we have shown by means of \lem that superconductivity in $\mathrm{La_{1.56}Sr_{0.44}CuO_4}/\mathrm{La_2CuO_4}$ superlattices originates from a charge transfer at the interfaces between LSCO and LCO, and thus preventing supercurrents to flow perpendicular to the SL ($c$-axis, $H_{\rm ext}^\|\,\|\, ab$-planes). The vortex state found is best described by a pancake-vortex model with negligible Josephson coupling and a superfluid density close to the one in optimally doped bulk LSCO.

\section*{Acknowledgments}
The \muSR experiments were fully performed at the S$\mu$S.
The work at BNL was supported by the U.S. Department of Energy Project MA-509-MACA.


\end{document}